\documentclass{aa}
\usepackage[varg]{txfonts}
\usepackage{graphicx}
\usepackage{float}
\graphicspath{ {./plots/} }
\usepackage{natbib}
\bibpunct{(}{)}{;}{a}{}{,}
\usepackage{multicol} 
\usepackage{array,multirow}
\usepackage{footnote}
\usepackage{url}
\usepackage[titletoc]{appendix}
\usepackage{tabularx}
\usepackage{soul}
\usepackage{babel}
\usepackage{xcolor}
\usepackage{hyperref}
\hypersetup{colorlinks, allcolors = {blue!50!black}}
\usepackage{cleveref}
\usepackage{orcidlink}
\begin{document} 

%\makeatletter\@openrightfalse\makeatother

\title{A Ring of Fire Orphan $\gamma$-Ray Flare in the Neutrino Candidate 3C 120}

\author{
E.~Traianou\inst{1}\orcidlink{0000-0002-1209-6500}
\and
G.~Bruni\inst{2}\orcidlink{0000-0002-5182-6289}
\and
J.~Rodi\inst{2}\orcidlink{0000-0003-2126-5908}
\and
G.~F.~Paraschos\inst{4,5,6}\orcidlink{0000-0001-6757-3098}
\and
S.~G.~Jorstad\inst{8}\orcidlink{0000-0001-9522-5453}
\and
A.~P.~Marscher\inst{8}\orcidlink{0000-0001-7396-3332}
\and
A.~L\"ahteenm\"aki\inst{5,7}\orcidlink{0000-0002-0393-0647}
\and
M.~Tornikoski\inst{5}\orcidlink{0000-0003-1249-6026}
\and
J.~Tammi\inst{5}\orcidlink{0000-0002-9164-2695}
\and
I.~Agudo\inst{3}\orcidlink{0000-0002-3777-6182}
}

\institute{
Interdisziplin\"ares Zentrum f\"ur Wissenschaftliches Rechnen (IWR), Universit\"at Heidelberg, Im Neuenheimer Feld 205, 69120 Heidelberg, Germany
\email{traianouthalia@gmail.com}
\and
INAF - Istituto di Astrofisica e Planetologia Spaziali, Via del Fosso del Cavaliere 100, I-00133 Rome, Italy
\and
Instituto de Astrof\'{\i}sica de Andaluc\'{\i}a (IAA-CSIC), Glorieta de la Astronom\'{\i}a s/n, 18008 Granada, Spain
\and
Finnish Centre for Astronomy with ESO, University of Turku, 20014 Turku, Finland
\and
Aalto University Mets\"ahovi Radio Observatory, Mets\"ahovintie 114, FI-02540 Kylm\"al\"a, Finland
\and
Max-Planck-Institut f\"ur Radioastronomie, Auf dem H\"ugel 69, D-53121 Bonn, Germany
\and
Aalto University Department of Electronics and Nanoengineering, P.O. Box 15500, FI-00076 Aalto, Finland
\and
Institute for Astrophysical Research, Boston University, 725 Commonwealth Avenue, Boston, MA 02215, USA
}

\date{Received [date] / Accepted [date]}

% \abstract{}{}{}{}{} 
% 5 {} token are mandatory

\abstract{We present 43\,GHz VLBI observations of the radio galaxy 3C~120 during its brightest $\gamma$-ray outburst (March 2018), recently associated with the IceCube neutrino alert IC-180213A. Despite reaching $L_\gamma = 3.7 \times 10^{44}$\,erg\,s$^{-1}$, contemporaneous X-ray monitoring from INTEGRAL/ISGRI, MAXI/GSC, and \textit{Swift}/XRT revealed no variability across 0.3-200\,keV, nor in B, V, R, and I band optical observations or 37 \& 235\,GHz observations, establishing an orphan flare. High-cadence VLBI imaging identified a new jet disturbance (N) propagating at $\beta_{\rm app} = (2.8 \pm 1.3)$ through quasi-stationary features C1-C3. The $\gamma$-ray peak coincided spatially and temporally with N crossing C3 ($r \sim 0.38$\,mas), where we measured a factor-of-5 increase in fractional polarization ($m = 16\%$) and $\Delta\chi \sim 24^\circ$ EVPA rotation, indicating localized magnetic field compression. The extreme Compton dominance ($L_\gamma / L_{\rm syn,blob} \approx 160$) is naturally explained by the Ring of Fire scenario, in which N ($\Gamma_{\rm blob} = 6$, $B_{\rm blob} = 0.023$\,G) inverse-Compton scatters synchrotron photons from C3, reproducing the observed $\gamma$-ray luminosity for physically reasonable parameters. We rule out alternative mechanisms: spine-sheath geometric boosting requires implausibly large viewing-angle swings, magnetic reconnection predicts excessive synchrotron emission, and blob-star collisions cannot explain the extended ($\sim$100-150\,day) period of elevated activity. Unlike the 2014-2015 orphan flares attributed to rapid spine reorientation near the BLR, the 2018 event represents a distinct physical mechanism, a propagating disturbance interacting with stationary jet structure at $\sim10\times$ the BLR radius.This work provides the first direct observational link between VLBI-resolved jet dynamics and orphan $\gamma$-ray emission in a radio galaxy.}


%\authorrunning{E. Traianou et.al.} 
\maketitle
%________________________________________________________________

\section{Introduction}
\label{sec:intro}

Orphan $\gamma$-ray flares are high-energy outbursts occurring without corresponding variability at other wavelengths, and appear to be intrinsically rare: among the hundreds of $\gamma$-ray blazars monitored by \textit{Fermi}-LAT over more than a decade, only a small number of well-documented orphan flares have been reported \citep{2004ApJ...601..151K,2015ApJ...804..111M,deJaeger2023}. This rarity makes each event particularly valuable, as orphan flares directly challenge standard single-zone models in which synchrotron and inverse-Compton emission are expected to vary together because they originate from the same electron population \citep{2019MNRAS.484.2067G,Sobacchi2021}.

Several theoretical frameworks have been proposed to account for orphan behavior. In the Ring of Fire model \citep{2015ApJ...804..111M}, a relativistic plasma blob propagates through a shocked jet sheath and inverse-Compton scatters the sheath’s synchrotron photons without producing a strong synchrotron flare itself. Other scenarios invoke structured leptonic jets with spatially separated emission regions \citep{2006ApJ...651..113K}, hadronic or lepto-hadronic processes generating cascades with distinct radiative signatures \citep{2005ApJ...621..176B}, or transient interactions between moving shocks and localized external photon fields such as stars or disk-driven outflows \citep{2010A&A...522A..97A,2016MNRAS.463L..26B}. Crucially, these models make different predictions for the location, timescale, and polarization properties of the emission.

The nearby radio galaxy 3C\,120 ($z=0.033$) offers an exceptional laboratory for testing these ideas. Although classified as a Fanaroff-Riley type~I radio galaxy \citep{2012ApJ...752...92A}, 3C\,120 exhibits pronounced blazar-like behavior on parsec scales. VLBI observations reveal multiple superluminal components with apparent speeds of $\beta_{\rm app}\sim$4-6 \citep{2000Sci...289.2317G,2001ApJ...561L.161G,2011ApJ...733...11G}, traceable to projected distances beyond 150\,pc from the core \citep{2001ApJ...556..756W}. At the same time, unlike blazars, 3C\,120 retains clear accretion-disk signatures: its X-ray spectrum shows Seyfert-like behavior with intensity-correlated spectral changes \citep{1991ApJ...368..138M}, and a prominent Fe~K$\alpha$ line at 6.4\,keV indicates that most of the X-ray emission originates close to the disk rather than in the jet \citep{2002Natur.417..625M}. This hybrid character makes 3C\,120 particularly well suited for disentangling disk and jet contributions.

Long-term multi-wavelength monitoring has established a tight disk-jet connection in 3C\,120. Pronounced X-ray dips are systematically followed, after a delay of $\sim$120~days, by the ejection of bright superluminal radio knots \citep[e.g.,][]{2009ApJ...704.1689C}, indicating that disturbances originate near the black hole and propagate outward before becoming visible at parsec scales. Since its detection by \textit{Fermi}-LAT in 2010 \citep{2010ApJ...720..912A}, 3C\,120 has exhibited episodic $\gamma$-ray activity, including major flares in 2012-2014 \citep{2015ApJ...808..162C} and extreme intraday orphan events in 2014-2015 \citep{2016MNRAS.458.2360J}. The latter were interpreted within a spine-sheath framework, where brief geometric changes in a fast inner spine produce strong Doppler boosting of external-Compton emission on BLR or torus photons, with little contemporaneous response at other wavelengths. Furthermore, VLBA polarimetric studies that reveal Faraday rotation gradients and systematic EVPA behavior further support a stratified jet structure guided by a helical magnetic field \citep{2001ApJ...556..756W,2008ApJ...681L..69G}.

Interest in 3C\,120 has increased further following the proposal by \citet{2025A&A...702A.129C} of an association between a strong $\gamma$-ray flare in early 2018 and the IceCube neutrino event IC-180213A. If confirmed, this would make 3C\,120 the first radio galaxy linked to high-energy neutrino emission, providing key insight into hadronic acceleration in mildly misaligned, structured jets.

Within this broader context, the $\gamma$-ray outburst of March 2018 stands out as the most luminous high-energy event ever observed from 3C\,120\href{https://fermi.gsfc.nasa.gov/ssc/data/access/lat/LightCurveRepository/index.html}{(Fermi-LAT Light Curve Repository)}. Comprehensive contemporaneous multi-wavelength 
monitoring showed no flaring behaviour at X-ray, optical, or radio wavelengths, firmly establishing the event as a clear orphan outburst. While orphan behavior was seen 
earlier from the source during the 2014-2015 activity \citep{2016MNRAS.458.2360J}, the 2018 outburst differs in its longer period of elevated activity ($\sim$100-150~days), its highest recorded peak luminosity, and dense VLBI coverage, which allows the jet response to be followed in detail.

In this paper, we present a multi-wavelength analysis of the 2018 outburst, combining \textit{Fermi}-LAT $\gamma$-ray observations, broad X-ray monitoring from INTEGRAL, MAXI, and \textit{Swift}, UV and optical monitoring including \textit{Swift}/UVOT UVW1 and 
B, V, R, and I band photometry from the Perkins telescope and CAFOS/2.2\,m, as well as radio observations at 37 \& 235\,GHz, including 43\,GHz VLBA imaging. Our aims are to characterize the outburst within the long-term $\gamma$-ray behavior of 3C\,120, identify and quantify contemporaneous parsec-scale jet changes, establish causal links between jet dynamics and high-energy emission, and assess the viability of different emission scenarios, including spine-sheath interactions, external Compton models, and the Ring of Fire framework. Throughout, we adopt a flat $\Lambda$CDM cosmology with $H_0 = 71$\,km\,s$^{-1}$\,Mpc$^{-1}$ \citep{2009ApJS..180..330K}, corresponding to a linear scale of 0.67\,pc\,mas$^{-1}$ at the distance of 3C\,120.

\section{Observations and data analysis}
\label{sec:obs}

\subsection{VLBI observations, imaging, and model-fitting}
\label{sec:vlbi_obs}

The 43\,GHz data presented in this work were obtained with the Very Long Baseline Array (VLBA) as part of the VLBA-BU-BLAZAR monitoring program\footnote{\url{https://www.bu.edu/blazars/BEAM-ME.html}} \citep{2017ApJ...846...98J,2022ApJS..260...12W}. We analyzed five epochs spanning December~24, 2017, to May~11, 2018 (MJD~58111-58249), providing dense temporal coverage of the $\gamma$-ray outburst. The calibrated data were imported into \texttt{Difmap} \citep{1997ASPC..125...77S}, using the CLEAN algorithm \citep{1974A&AS...15..417H} with iterative phase and amplitude self-calibration to produce total intensity and linear polarization maps.

\begin{figure}
    \centering
    \includegraphics[scale=0.085]{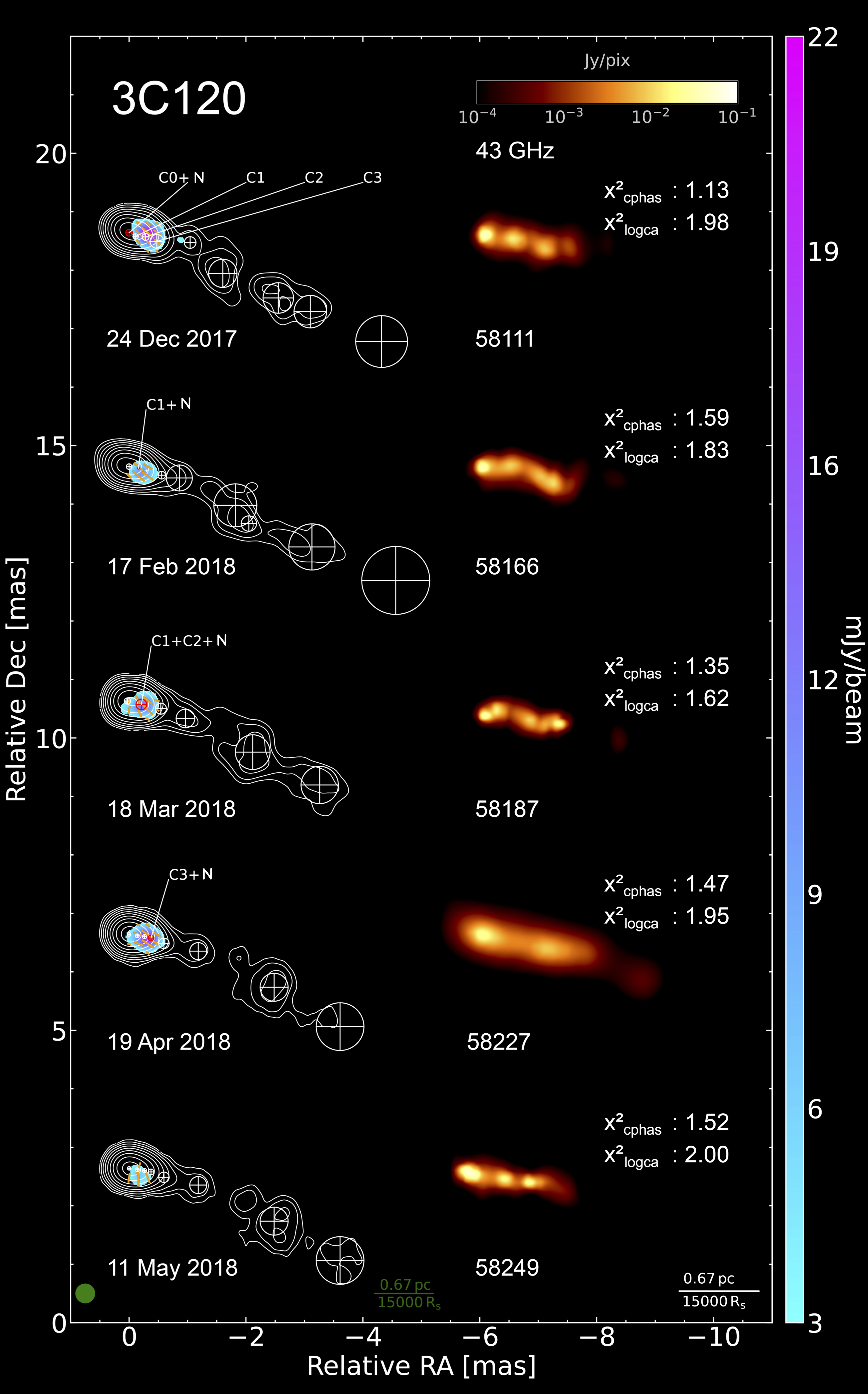}
\caption{43\,GHz VLBA images of 3C\,120 during the 2017-2018 monitoring sequence. 
Left: Total intensity (Stokes I) CLEAN maps shown as contours at levels of $[0.25, 0.5, 1, 2, 4, 8, 16, 32, 64]\%$ of the Stokes I peak ($1.5$\,Jy\,beam$^{-1}$), whereas orange sticks indicate the EVPA. Regions without significant polarized signal are blanked. Circular Gaussian components from the visibility-domain model fits are overlaid and labeled to identify the core and the inner quasi-stationary features (C1-C3), as well as epochs where the emerging disturbance is unresolved from these structures 
(e.g., C0+N, C1+C2+N, C3+N). The common restoring beam (FWHM $= 0.32$\,mas, geometrical mean) is shown in green in the lower left, and the maps are aligned in relative right ascension and declination (mas), a linear scale bar is provided for reference. Right: Corresponding regularized maximum-likelihood reconstructions obtained with \texttt{eht-imaging}, the observing epoch (MJD) and the reduced $\chi^{2}$ values for closure phases and logarithmic closure amplitudes are indicated in each panel.}
\label{fig:main}
\end{figure}

In parallel, we reconstructed the 43\,GHz images using a regularized maximum likelihood (RML) approach implemented in the \texttt{eht-imaging} library \citep{Chael_2016,2018ApJ...857...23C}. The fully calibrated visibilities were imaged over a $2\times2$\,mas field of view on a $300\times300$ pixel grid. We minimized the standard \texttt{ehtim} objective function combining data-fidelity and regularization terms, adopting the same set of regularizers and relative entropy, total variation, and $\ell_1$ sparsity, as in \citet{2019ApJ...875L...4E}. The regularization weights were optimized via a small grid search to achieve reduced $\chi^2 \simeq 1$ for closure phases and logarithmic closure amplitudes. The resulting images were subsequently self-calibrated in phase and amplitude to improve dynamic range. The resulting images are shown in Figure~\ref{fig:main}.

Model fitting was performed directly in the visibility domain using circular Gaussian components with the \texttt{MODELFIT} routine in \texttt{Difmap}. Individual components were identified and tracked across epochs based on their flux densities, separations from the core, and position angles, following the approach of \citet{2024A&A...682A.154T}. Parameter uncertainties were estimated from the local signal-to-noise ratio, with flux density uncertainties conservatively set to 10\% \citep{2009AJ....138.1874L}. The full set of model-fitting results is summarized in Table~\ref{tab:model_fit}.

\begin{table*}
\centering
\caption{Model–fitting parameters of the VLBI components.}
\begin{tabular}{@{}lcccccccccc@{}}
\hline\hline
\noalign{\smallskip}
Comp. & Epoch & $S$ & $\Delta S$ & $r$ & $\Delta r$ & $\theta$ & $\Delta\theta$ & FWHM & $\Delta{\rm FWHM}$ \\
 &  & (Jy) & (Jy) & (mas) & (mas) & ($^\circ$) & ($^\circ$) & (mas) & (mas) \\
(1) & (2) & (3) & (4) & (5) & (6) & (7) & (8) & (9) & (10) \\
\noalign{\smallskip}
\hline
\noalign{\smallskip}
C0+N & 2018.0 & 0.46 & 0.05 & 0.00 & 0.00 &   0 &  0 & 0.11 & 0.01 \\
\multirow{4}{*}{C0}
 & 2018.1 & 0.55 & 0.06 & 0.00 & 0.00 &   0 &  0 & 0.02 & 0.00 \\
 & 2018.2 & 0.89 & 0.09 & 0.00 & 0.00 &   0 &  0 & 0.11 & 0.01 \\
 & 2018.3 & 1.20 & 0.12 & 0.00 & 0.00 &   0 &  0 & 0.06 & 0.01 \\
 & 2018.4 & 0.87 & 0.09 & 0.00 & 0.00 &   0 &  0 & 0.05 & 0.01 \\
\hline
C1        & 2018.0 & 0.34 & 0.03 & 0.13 & 0.05 & -110 &  0 & 0.09 & 0.01 \\
C1+N    & 2018.1 & 0.40 & 0.04 & 0.16 & 0.03 & -100 &  0 & 0.03 & 0.00 \\
C1+C2+N & 2018.2 & 0.81 & 0.08 & 0.25 & 0.04 & -104 & 10 & 0.18 & 0.02 \\
\multirow{2}{*}{C1}
 & 2018.3 & 0.40 & 0.04 & 0.14 & 0.04 & -100 &  4 & 0.07 & 0.01 \\
 & 2018.4 & 0.53 & 0.05 & 0.13 & 0.03 & -100 &  4 & 0.08 & 0.01 \\
\hline
\multirow{4}{*}{C2}
 & 2018.0 & 0.37 & 0.04 & 0.29 & 0.06 & -104 & 9 & 0.12 & 0.01 \\
 & 2018.1 & 0.55 & 0.06 & 0.31 & 0.06 & -103 &  2 & 0.10 & 0.01 \\
 & 2018.3 & 0.34 & 0.03 & 0.27 & 0.04 & -98  &  7 & 0.07 & 0.01 \\
 & 2018.4 & 0.40 & 0.04 & 0.27 & 0.03 & -102 &  7 & 0.09 & 0.01 \\
\hline
\multirow{3}{*}{C3}
 & 2018.0 & 0.14 & 0.01 & 0.50 & 0.06 & -107 & 10 & 0.23 & 0.02 \\
 & 2018.1 & 0.05 & 0.01 & 0.58 & 0.06 & -105 & 10 & 0.13 & 0.01 \\
 & 2018.2 & 0.03 & 0.003 & 0.59 & 0.04 & -103 &  4 & 0.19 & 0.02 \\
C3+N & 2018.3 & 0.10 & 0.01 & 0.38 & 0.04 & -99  &  5 & 0.10 & 0.01 \\
    & 2018.4 & 0.09 & 0.01 & 0.42 & 0.03 & -99  &  4 & 0.15 & 0.01 \\
\hline
\multirow{5}{*}{C4}
 & 2018.0 & 0.02 & 0.002 & 1.10 & 0.06 & -99  &  4 & 0.20 & 0.02 \\
 & 2018.1 & 0.03 & 0.003 & 0.90 & 0.08 & -103 &  6 & 0.44 & 0.04 \\
 & 2018.2 & 0.04 & 0.004 & 1.00 & 0.06 & -107 &  3 & 0.32 & 0.03 \\
 & 2018.3 & 0.03 & 0.003 & 0.62 & 0.04 & -104 &  4 & 0.29 & 0.03 \\
 & 2018.4 & 0.03 & 0.003 & 0.69 & 0.04 & -104 &  3 & 0.23 & 0.02 \\
\hline
\multirow{5}{*}{C5}
 & 2018.0 & 0.08 & 0.01 & 1.80 & 0.09 & -113 &  3 & 0.48 & 0.05 \\
 & 2018.1 & 0.05 & 0.01 & 1.90 & 0.12 & -110 &  4 & 0.73 & 0.07 \\
 & 2018.2 & 0.11 & 0.01 & 2.30 & 0.10 & -112 &  2 & 0.59 & 0.06 \\
 & 2018.3 & 0.03 & 0.00 & 1.20 & 0.06 & -104 &  3 & 0.29 & 0.03 \\
 & 2018.4 & 0.03 & 0.00 & 1.30 & 0.07 & -104 &  3 & 0.33 & 0.03 \\
\hline
\multirow{4}{*}{C6}
 & 2018.0 & 0.04 & 0.004 & 2.79 & 0.09 & -114 &  2 & 0.52 & 0.05 \\
 & 2018.1 & 0.03 & 0.003 & 2.27 & 0.07 & -115 &  2 & 0.26 & 0.03 \\
 & 2018.3 & 0.07 & 0.01 & 2.60 & 0.08 & -110 &  2 & 0.48 & 0.05 \\
 & 2018.4 & 0.02 & 0.002 & 2.10 & 0.06 & -105 &  2 & 0.30 & 0.03 \\
\hline
\multirow{5}{*}{C7}
 & 2018.0 & 0.04 & 0.004 & 3.40 & 0.10 & -114 &  2 & 0.56 & 0.06 \\
 & 2018.1 & 0.06 & 0.01 & 3.40 & 0.12 & -114 &  2 & 0.79 & 0.08 \\
 & 2018.2 & 0.06 & 0.01 & 3.60 & 0.09 & -114 &  1 & 0.64 & 0.06 \\
 & 2018.3 & 0.03 & 0.003 & 3.90 & 0.08 & -114 &  1 & 0.82 & 0.08 \\
 & 2018.4 & 0.07 & 0.01 & 2.80 & 0.06 & -111 &  1 & 0.54 & 0.05 \\
\hline
\multirow{3}{*}{C8}
 & 2018.0 & 0.02 & 0.002 & 4.70 & 0.11 & -113 &  1 & 0.89 & 0.09 \\
 & 2018.1 & 0.01 & 0.001 & 4.90 & 0.15 & -113 &  2 & 1.20 & 0.12 \\
 & 2018.4 & 0.04 & 0.004 & 4.00 & 0.09 & -114 &  1 & 0.80 & 0.08 \\
\hline
\end{tabular}
\tablefoot{Columns: (1) model component ID, (2) observing epoch in decimal years, (3) flux density, (4) flux density uncertainty, (5) angular separation from the core, (6) uncertainty in separation, (7) position angle, (8) uncertainty in position angle, (9) FWHM of the Gaussian component, (10) FWHM uncertainty. The component labeled C0+N at 2018.0 reflects the unresolved blend of the VLBI core with the emerging new disturbance N. Similarly, labels such as C1+N, C1+C2+N, and C3+N indicate epochs when the propagating disturbance is superimposed on stationary jet features and cannot be resolved into separate Gaussian components.}
\label{tab:model_fit}
\end{table*}

\subsection{Multiwavelength data}
\label{sec:multi}

The multi-wavelength data used in this work are summarized in Fig.~\ref{fig:mwl_lc}, which highlights the 2018 March 30 $\gamma$-ray outburst together with the contemporaneous coverage across X-ray, UV, optical, and radio bands. 

\subsubsection{\textit{Fermi}/LAT $\gamma$-ray data}
$\gamma$-ray data were obtained from the public \textit{Fermi}/LAT archive, covering the 0.1-100\,GeV energy range. To place the 2018 outburst in a broader context, we constructed a monthly binned light curve over 2017-2018. The analysis followed standard LAT procedures using the \texttt{Fermitools} software package together with the latest instrument response functions. We applied the recommended data-quality selections to retain only high-confidence events and performed a binned maximum-likelihood analysis to derive flux measurements. The source spectrum was modeled with a simple power law. Statistical uncertainties were taken from the likelihood fit, while systematic uncertainties were estimated via bootstrap resampling. To further characterize the temporal structure of the outburst, we  additionally computed $\gamma$-ray light curves with 7-day and 15-day  bins. These confirm that the activity peaks near MJD~$\sim$58200 but do not allow finer localization of the exact peak date given the  available photon statistics.

\subsubsection{X-ray data and variability analysis}
Statistical analysis of X-ray variability was performed using the normalized excess variance and fractional root-mean-square variability methods \citep{2003MNRAS.345.1271V}. To assess whether the March 2018 $\gamma$-ray outburst has a contemporaneous X-ray counterpart, we examined X-ray monitoring data from multiple instruments spanning soft to hard X-rays. For the hard X-ray band, we used the 1-day \textit{Swift}/BAT 15-50\,keV light curve\footnote{\url{https://swift.gsfc.nasa.gov/results/transients/}} and \textit{INTEGRAL}/ISGRI light curves in four energy bands (20-40, 40-60, 60-100, and 100-200\,keV) on a 1-day timescale. The ISGRI data were analyzed with the standard INTEGRAL Offline Science Analysis (OSA) v11.2 software\footnote{\url{https://www.isdc.unige.ch/integral/analysis\#Software}}. These time series correspond to the products used in a previous publication and were provided by J. Rodi. 
For the soft X-ray band, we used the public MAXI/GSC 1-day light curve data in 2-20\,keV, including the 2-4, 4-10, and 10-20\,keV sub-bands (see MAXI archive for the column definitions). In addition, we include \textit{Swift}/XRT 0.3-10\,keV pointed observations overlapping with the outburst window when available; XRT count rates (and fluxes where applicable) are derived using standard \textit{Swift} tools with an absorbed power-law conversion appropriate for 3C\,120.

\subsubsection{Optical data}
We have performed optical polarimetric R band and photometric BVRI observations of 3C\,120 using the 1.83\,m Perkins telescope (Flagstaff, AZ, USA) equipped with the PRISM camera\footnote{\url{https://www.bu.edu/prism/specs.htm}}. We employed a differential photometry method to measure BVRI magnitudes with the comparison stars 1, 2, and 3 from \cite{2005Ap.....48..304D}. The polarimetric observations were carried out with a rotating polaroid (POL-HN38). A circular aperture with a radius of $7.5\arcsec$ was applied to measure the core of 3C\,120, stars in the field, and the sky background. The images were corrected for bias and flat field. Each polarization observation involved 3 series of Stokes I, Q, and U measurements, with a series consisting of four measurements at polaroid position angles of 0$^\circ$, $90^\circ$, 45$^\circ$, and 135$^\circ$. Values of Q, and U parameters were averaged over the series to calculate the degree of polarization and position angle of polarization and their uncertainties. Since the camera has a wide field of view (14'$\times$14'), both interstellar and instrumental polarization corrections were performed using field stars, assuming that they are intrinsically unpolarized. We utilized unpolarized calibration stars from \cite{1992AJ....104.1563S} to check the instrumental polarization, which is usually within 0.2\%, and polarized stars from the same paper to calibrate the polarization position angle. Additional R-band photometric observations of 3C\,120 were obtained with the CAFOS instrument at the 2.2\,m Telescope of the Calar Alto Observatory\footnote{Calar Alto data was acquired as part of the MAPCAT project: \url{http://www.iaa.es/~iagudo/_iagudo/MAPCAT.html}} (Almer\'ia, Spain). The CAFOS data were reduced in the same way as those from Perkins, and they cover the period surrounding the $\gamma$-ray outburst. They include an observation at MJD\,58205.8 ($R = 14.16 \pm 0.06$\,mag), just one day before the $\gamma$-ray peak. A single \textit{Swift}/UVOT UVW1 observation at MJD\,58196 ($m_{\rm UVW1} = 14.53 \pm 0.02$\,mag, AB system) was also retrieved from the public \textit{Swift} archive.

\subsubsection{Radio data}
Radio variability data contemporaneous with the VLBI observations were compiled from mm-cm monitoring programs. Measurements at 235\,GHz were obtained with the 8-element Submillimeter Array (SMA; \citet{2007ASPC..375..234G}), while the 37\,GHz light curve was taken from the long-term monitoring of the Aalto University Mets{\"a}hovi Radio Observatory (MRO)\footnote{\url{https://www.aalto.fi/en/metsahovi-radio-observatory}}, which operates a 14-m telescope providing near-daily radio observations of hundreds of AGN.

\begin{figure*}
    \centering
    \includegraphics[scale=0.23]{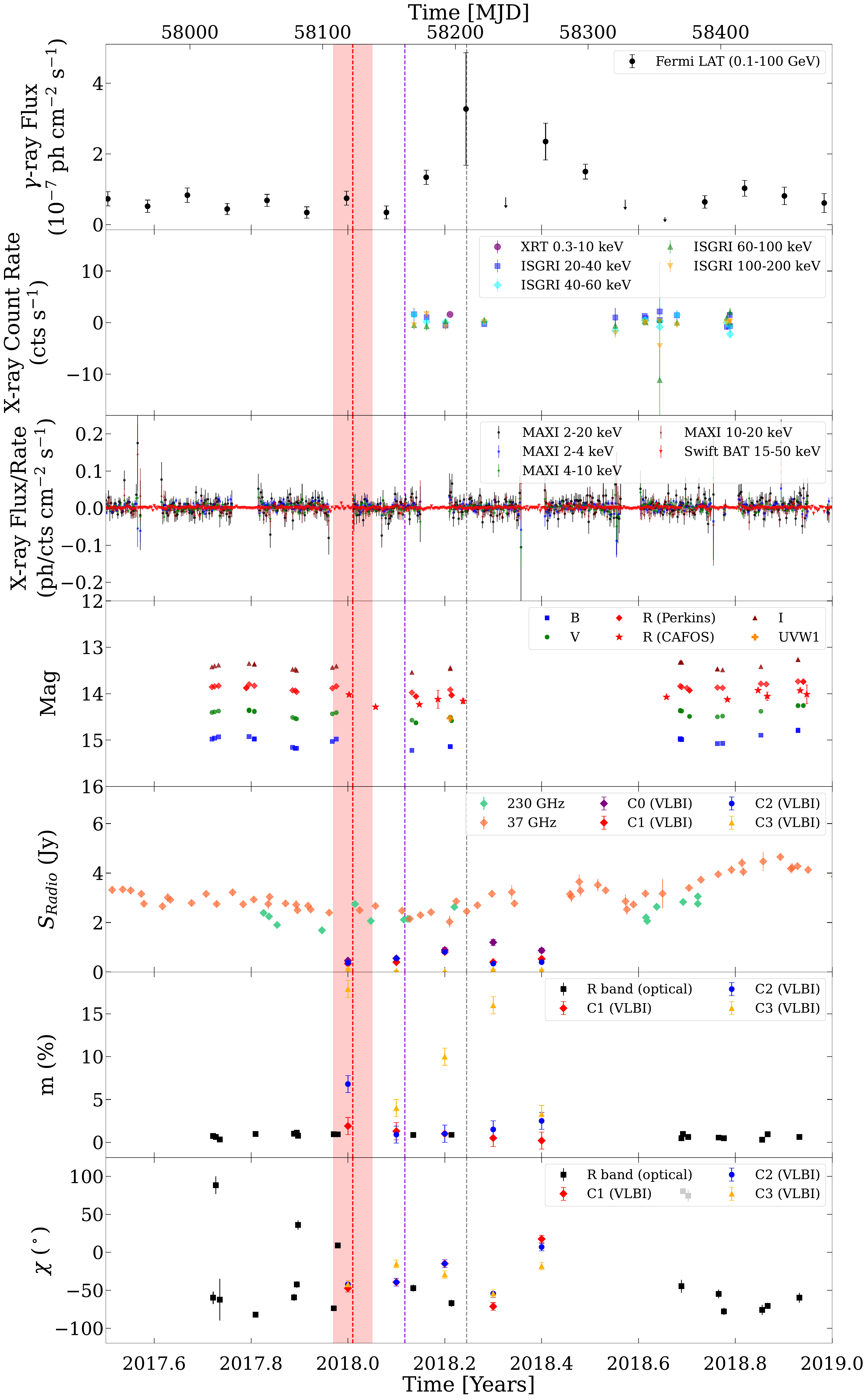}
\caption{Multi-wavelength light curve of 3C\,120 (2017--2019). Vertical lines mark: the estimated ejection time of N (red dashed, with red shaded uncertainty), the $\gamma$-ray peak (grey dashed, MJD\,58208), and the IceCube neutrino IC-180213A arrival (dark violet dashed, MJD\,58162). From top to bottom: \textit{Fermi}-LAT $\gamma$-ray flux (0.1-100\,GeV), \textit{Swift}/XRT (purple circles) and INTEGRAL/ISGRI in four bands: 20-40\,keV (blue squares), 40-60\,keV (cyan diamonds), 60-100\,keV (green triangles), 100-200\,keV (orange triangles), MAXI/GSC (black circles) and \textit{Swift}/BAT 15-50\,keV (red triangles), optical B (blue squares), V (green circles), R (red diamonds/stars from Perkins/CAFOS), I (dark red triangles), and \textit{Swift}/UVOT UVW1 at MJD\,58196 (orange plus, AB mag), radio flux densities from SMA 230\,GHz (green diamonds), 37\,GHz (orange diamonds), and VLBI components C0 (purple), C1 (red), C2 (blue), C3 (yellow), fractional linear polarization for optical R-band (black squares) and VLBI components C1, C2, C3, EVPA for the same dataset. X-ray bands remain quiescent throughout, while enhanced polarization in C3 coincides with the $\gamma$-ray peak.}
\label{fig:mwl_lc}
\end{figure*}

\section{Results}
\label{sec:results}

\subsection{VLBI jet structural evolution}
\label{sec:vlbi_results}

At 43\,GHz the source is characterized by a bright, compact core, namely C0, at the upstream end of a one-sided jet extending toward the southwest. The inner jet segment is populated by a sequence of quasi-stationary features (C1, C2, and C3) at projected separations of $\sim$0.1-0.6\,mas, followed by a train of more diffuse downstream components (C4-C8) extending to several mas from the core (Table~\ref{tab:model_fit}, and Figure~\ref{fig:main}). Superposed on this stationary pattern, we report the emergence of a new feature (hereafter N), where its presence is confirmed from systematic, and transient changes in the flux density, size, and polarization of C1, C2, and C3. 

Specifically, in 2018.0, N is blended with the VLBI core, but by 2018.1, a clear enhancement appears in C1 (C1+N), with a flux density increase of $\sim$20\% relative to the previous epoch (0.1\,Jy). At 2018.2 the inner jet becomes both brighter and broader, with the emission represented by C1+C2+N ($S=0.81\pm0.08$\,Jy), nearly doubling the flux density. In subsequent epochs the flux enhancement shifts farther downstream: at 2018.3-2018.4 the disturbance is detected as C3+N, while C1 returns to its pre-flare parameters. Anchoring the kinematics to the epochs of peak flux enhancement, we associate C1+C2+N at 2018.2 ($r=(0.25\pm0.04)$\,mas) and C3+N at 2018.3 ($r=(0.38\pm0.04)$\,mas). Following the approach of \citet{2020A&A...634A.112T,2025A&A...695L...3P}, this yields a pattern speed $\mu = (1.30\pm0.57)$\,mas\,yr$^{-1}$ (corresponding to $\beta_{\rm app}=(2.8\pm1.3)$) and, by linear back-extrapolation to $r=0$, a core-separation epoch $t_{\rm ej}=(2018.01\pm0.08)$\,years, consistent with N being blended with the VLBI core at 2018.0. 

We note that this feature N is kinematically distinct from the faster moving component C15 reported by \citet{2022ApJS..260...12W}, which has $\mu = (2.59\pm0.04)$\,mas\,yr$^{-1}$ ($\beta_{\rm app}=(5.75\pm0.09)$) and was ejected at $t_{\rm ej}=(2017.83\pm0.08)$. At our first epoch (2018.0), C15 appears to be crossing through the quasi-stationary feature C3 ($r\approx$0.45-0.50\,mas) before continuing downstream to $r\approx0.71$\,mas by 2018.1 and $r\approx1.23$\,mas by 2018.3. The presence of both the fast-moving C15 and the slower disturbance N indicates heightened jet activity during the 2017-2018 period.

The \texttt{eht-imaging} reconstructions, with effective resolution improved by up to a factor of $\sim$3 compared to the CLEAN images, provide a sharper view of the inner $\sim$0.6\,mas and corroborate the model fitting results. In the first epoch, the innermost structure separates cleanly into the VLBI core C0 and stationary features C1-C3. By the next epoch, the inner jet shows morphological changes consistent with either a local position-angle swing or selective brightening of a different streamline as a disturbance propagates through a stratified flow. This effect strengthens in the following epoch, where two distinct bends become apparent, one within a few tens of $\mu$as from the core and another immediately upstream of C3, while the innermost emission appears as C1 and C2 blended with N. In the epoch just after the $\gamma$-ray outburst, the core and C3 region dominate the emission, indicating concentrated dissipation in the inner jet; in the final epoch the core and C3 appear elongated toward the southwest, plausibly tracing N as it exits the C3 region.

The polarization behavior follows the same picture. In the inner jet, C1 shows a clear polarization response during the disturbance passage: its electric vector position angle (EVPA, $\chi = \tfrac{1}{2}\arctan(U/Q)$) rotates from $\chi\simeq-39^\circ$ at 2018.1 to $\chi\simeq-15^\circ$ at 2018.2 ($\sim24^\circ$), followed by a rapid swing to $\chi\simeq-71^\circ$ at 2018.3. Together with contemporaneous variations in linear polarization flux $P$ ($P = \sqrt{Q^2 + U^2}$, and fractional polarization $m$ ($m = P/S \times 100\%$ \cite{2025A&A...700A..16T}), these EVPA rotations indicate a disturbance/shock compressing and partially ordering the local magnetic field as it crosses a standing feature, with detailed EVPA behavior likely modulated by opacity and blending effects. Outside the brief blending phases, C2 varies only modestly, with $m$ remaining at the few-percent level. In contrast, C3 exhibits a pronounced polarization enhancement during the downstream phase: $P$ reaches $\sim16$\,mJy at 2018.3 and fractional polarization rises to $\sim16\%$, pointing to a highly ordered field when the disturbance interacts with this stationary feature. Over the outburst window we also measure an overall EVPA shift of $\sim20^\circ$ toward the jet direction, consistent with the shocked region developing a predominantly transverse projected magnetic field (perpendicular to the jet axis), as expected from compression of a tangled field and/or amplification of the transverse component in a recollimation/shock zone. The polarization properties and their temporal evolution for all VLBI components are summarized in Table~\ref{tab:pol_components} and Figure~\ref{fig:mwl_lc}.

\begin{table}
\centering
\caption{Linear polarization properties of VLBI model components at 43\,GHz.}
\begin{tabular}{@{}cccccc@{}}
\hline\hline
\noalign{\smallskip}
Comp. & Epoch & $P$ & $\Delta P$ & $\chi$ & $m$ \\
 &  & (mJy) & (mJy) & ($^\circ$) & (\%) \\
(1) & (2) & (3) & (4) & (5) & (6) \\
\noalign{\smallskip}
\hline
\noalign{\smallskip}
\multirow{5}{*}{C1}
 & 2018.0 & 6.5  & 0.8 & -47.2 & 1.9 \\
 & 2018.1 & 5.0  & 0.6 & -39.3 & 1.3 \\
 & 2018.2 & 8.0  & 2.0   & -14.9 & 1.0 \\
 & 2018.3 & 2.0  & 0.7 & -71.2 & 0.5 \\
 & 2018.4 & 1.0  & 0.4 & 17.5  & 0.2 \\
\hline
\multirow{4}{*}{C2}
 & 2018.0 & 25.0 & 3.0   & -42.1 & 6.8 \\
 & 2018.1 & 5.1  & 0.6 & -39.3 & 0.9 \\
 & 2018.3 & 5.0  & 0.6 & -54.3 & 1.5 \\
 & 2018.4 & 10.0 & 1.0   & 7.0   & 2.5 \\
\hline
\multirow{5}{*}{C3}
 & 2018.0 & 25.0 & 3.0   & -42.1 & 17.9 \\
 & 2018.1 & 2.0  & 0.3 & -15.2 & 4.0 \\
 & 2018.2 & 3.0  & 2.0   & -29.2 & 10.0 \\
 & 2018.3 & 16.0 & 2.0   & -54.2 & 16.0 \\
 & 2018.4 & 3.0  & 0.7 & -18.1 & 3.3 \\
\hline
\end{tabular}
\tablefoot{Columns: (1) model component, (2) observing epoch in decimal years, (3) polarized flux density, (4) uncertainty in polarized flux, (5) EVPA (6) fractional linear polarization. A systematic uncertainty of $5^\circ$ is adopted for $\chi$ based on absolute calibration accuracy, and 1\% for $m$ accounting for instrumental polarization leakage.}
\label{tab:pol_components}
\end{table}

\subsection{Multi-wavelength correlations}
\label{sec:correlations}

Figure~\ref{fig:mwl_lc} reveals the striking orphan nature of the March 2018 $\gamma$-ray outburst (peak MJD 58208, red line). While Fermi-LAT detects a dramatic $\gamma$-ray outburst, X-ray bands from soft (MAXI/GSC, Swift/XRT) to hard (BAT, INTEGRAL/ISGRI) remain completely flat. Optical photometry shows only gentle evolution, and mm/radio fluxes (230\,GHz, 37\,GHz) vary modestly. Crucially, an R-band CAFOS observation at MJD\,58205.8, just one day before the $\gamma$-ray peak, shows the source at $R = 14.16 \pm 0.06$\,mag, fully consistent with the quiescent baseline seen throughout the campaign. A \textit{Swift}/UVOT UVW1 observation at MJD\,58196 similarly shows no UV enhancement. Together these data strongly support the orphan interpretation, although a coincident or slightly delayed optical flare cannot be completely excluded given the limited sampling.

The smoking gun appears in our VLBI data: sequential brightening sweeps through C1 to C2 to C3, with the latter peaking exactly at 2018.3 when N is crossing it. Polarization measurements reveal C1 EVPA swings of 24$^{\circ}$ during passage, where C3 polarization spikes to 16\% right at the $\gamma$-ray maximum, with the jet disturbance N as the most plausible trigger of this high-energy event.

\section{Discussion}
\label{sec:discussion}

\subsection{Physical mechanisms: discriminating between emission scenarios}
\label{sec:mechanisms}

The orphan nature of the 2018 outburst, combined with our VLBI observations of jet structural evolution, allows us to test competing theoretical frameworks for high-energy emission in structured jets. We evaluate five distinct scenarios: (i) the Ring of Fire external Compton model, (ii) Geometric Doppler boosting, (iii) magnetic reconnection in a stratified flow, (iv) proton synchrotron / lepto-hadronic cascades, and (v) transient jet-star (or dense cloud) interactions that can locally enhance target photon fields and trigger short-lived high-energy outbursts.

\subsubsection{Ring of Fire model}
\label{sec:ring_of_fire}

The relative bulk motion between spine and sheath boosts the seed photon density in the blob frame by a factor $\sim\Gamma^2_{\rm rel}$, where $\Gamma_{\rm rel} = \Gamma_{\rm blob}\Gamma_{\rm sheath}(1 - \beta_{\rm blob}\beta_{\rm sheath})$ is the Lorentz factor of the blob measured in the sheath rest frame, providing an amplified external photon field for inverse-Compton scattering. For the simplified geometry adopted below, where C3 is treated as quasi-stationary in the AGN frame, we take $\Gamma_{\rm rel}\approx\Gamma_{\rm blob}=6$, corresponding to a seed-photon density enhancement of $\sim\Gamma_{\rm rel}^2\approx36$ in the blob frame.

To test the applicability of this model to the 2018 orphan outburst, we identify the quasi-stationary feature C3 at $r\sim0.38$\,mas as the seed-photon source. We adopt the following parameters based on our observations and the literature: C3 FWHM$_{\rm C3}=0.19\pm0.02$\,mas from quiescent-state measurements (2018.2), C3 quiescent flux $S_{\rm C3}=0.03$\,Jy at 43\,GHz, spectral index $\alpha=-0.5$ for optically thin synchrotron emission, C3 Doppler factor $\delta_{\rm C3}=3\pm1$, corresponding to the jet bulk flow with $\Gamma_{\rm jet}\sim5$ \citep{2015ApJ...808..162C}, blob Lorentz factor $\Gamma_{\rm blob}=6\pm1$ (adopting a conservative value within the MOJAVE-derived range of 5.4-8.1 from \citealt{2019ApJ...874...43L}), jet viewing angle $\theta_{\rm obs}=(15\pm3)^\circ$ \citep{2005AJ....130.1418J}, yielding blob Doppler factor $\delta_{\rm blob}=3.5\pm0.5$, blob size FWHM$_{\rm blob}\sim$ FWHM$_{\rm C3}$ (as N is unresolved, we adopt C3 size as an upper limit), and electron energy distribution parameters $\gamma_{\rm min}=1$, $\gamma_{\rm max}=2\times10^4$, and power-law index $s=2.4$ from \citet{2016MNRAS.458.2360J}.

Integrating the C3 synchrotron spectrum from 1\,GHz to 10\,THz \citep{1983ApJ...264..296M,2005AJ....130.1418J}, we obtain an intrinsic synchrotron luminosity $L_{\rm syn,C3}=6.4\times10^{40}$\,erg\,s$^{-1}$. 
The seed-photon energy density in the blob frame is calculated as $u'_{\rm ph} = \Gamma^2_{\rm blob}(1+\beta_{\rm blob})^2 \times L_{\rm syn,C3}/(4\pi R^2_{\rm C3}c) = 1.7\times10^{-4}$\,erg\,cm$^{-3}$, where $\beta_{\rm blob} = \sqrt{1 - 1/\Gamma^2_{\rm blob}}$ is the blob velocity in units of $c$. This expression assumes a head-on collision geometry between the blob and the C3 photon field, since C3 is quasi-stationary in the AGN frame, $\Gamma_{\rm rel} \approx \Gamma_{\rm blob}$.

The magnetic field in the blob is constrained by requiring the model to match the observed $\gamma$-ray luminosity. The peak $\gamma$-ray flux at MJD~58208 (epoch 2018.2), $(3.27\pm1.59)\times10^{-7}$\,ph\,cm$^{-2}$\,s$^{-1}$ in the 0.1-100\,GeV band, corresponds to an energy flux $(1.52\pm0.74)\times10^{-10}$\,erg\,cm$^{-2}$\,s$^{-1}$ (assuming a photon index $\Gamma_\gamma=2.5$) and a luminosity $L_\gamma=3.7\times10^{44}$\,erg\,s$^{-1}$. In the Thomson regime, the inverse-Compton to synchrotron luminosity ratio is $L_{\rm IC}/L_{\rm syn,blob}=u'_{\rm ph}/u'_{\rm B}$, where $u'_{\rm B}=B^2/(8\pi)$ is the magnetic energy density. With the blob's intrinsic synchrotron luminosity $L_{\rm syn,blob}=1.0\times10^{41}$\,erg\,s$^{-1}$ (from $S_{\rm blob}=0.07$\,Jy), we find that a magnetic field $B_{\rm blob}=(0.023\pm0.005)$\,G reproduces the observed $\gamma$-ray luminosity at the order-of-magnitude level, with agreement at the level of a few per cent for the adopted parameter set. We note that this estimate is based on a simplified analytic calculation rather than a full SED or time-dependent model, and is intended to demonstrate the plausibility of the scenario. This value is consistent with typical magnetic fields in relativistic blobs (0.01-0.1\,G; \citealt{2015ApJ...804..111M}), and is a factor $\sim7$ weaker than the underlying 3C~120 jet magnetic field at 0.1\,pc \citep{2016MNRAS.458.2360J}, which is physically reasonable for transient, dissipative structures undergoing magnetic reconnection or turbulent dissipation.

Crucially, the $\gamma$-ray outburst peaks at epoch 2018.2, coincident within uncertainties with the passage of N through C3 (epoch 2018.3 from model-fitting), providing direct observational support for inverse-Compton scattering of C3 synchrotron photons. Polarimetric signatures are consistent with this picture as well. An EVPA rotation of $\Delta\chi\sim24^\circ$ in C1 and a strong increase in fractional polarization in C3 ($m\sim16\%$ at epoch 2018.3) indicate magnetic-field compression as the disturbance crosses a standing shock, as expected for blob-stationary-feature interaction \citep{2007AJ....134..799J}.

The estimated C3 crossing time, $\Delta t\approx28$\,days (from C3 ${\rm FWHM}\sim0.1$\,mas and $\mu_{\rm N}=1.30$\,mas\,yr$^{-1}$), is shorter than the observed $\gamma$-ray activity, which extends over $\gtrsim$100-150~days. This indicates that the emission cannot arise from a single, localized interaction alone. 

A more plausible explanation is that the disturbance propagates through an extended region of enhanced seed-photon density, either due to the finite spatial extent of the shocked structure associated with C3 or through successive interactions with multiple quasi-stationary features (C1-C3) distributed over $\Delta r\approx0.25$\,mas. In this picture, the C3 interaction produces the dominant peak, while upstream interactions contribute to the extended activity.

Inverse-Compton cooling in the blob frame may contribute to smoothing the emission on shorter timescales, but cannot by itself account for the prolonged duration once the disturbance moves beyond the main seed-photon region.

Finally, although the VLBI core brightens significantly during the outburst interval, its lack of detectable polarization at any epoch, consistent with Faraday depolarization commonly observed in AGN cores \citep{2007AJ....134..799J,2008ApJ...681L..69G}, prevents us from directly diagnosing dissipation activity there. Also, while the compact core provides higher seed photon density for inverse-Compton scattering, the temporal and spatial coincidence of the $\gamma$-ray peak with N's passage through C3 (within $\sim10$ days and $\sim0.05$\,mas), together with the polarization maximum in C3 ($m = 16\%$), identifies this downstream region as the observationally confirmed Ring of Fire interaction site responsible for the orphan $\gamma$-ray outburst.

\subsubsection{Geometric Doppler Boosting in a Spine-Sheath Jet}
\label{sec:spine_sheath_doppler}

An alternative structured-jet scenario invokes rapid viewing-angle changes in a fast spine ($\Gamma_{\rm spine} \sim 10$-$20$) embedded within a slower sheath ($\Gamma_{\rm sheath} \sim 2$-5), producing large Doppler factor variations \citep{2005A&A...432..401G,2016MNRAS.458.2360J}. \citet{2016MNRAS.458.2360J} proposed this mechanism to explain the 2014-2015 sub-day orphan flares in 3C~120 via external Compton scattering of BLR photons near the VLBI core.

We test whether geometric Doppler boosting can explain the 2018 outburst. Adopting $\Gamma_{\rm blob}=6$ and $\theta_{\rm obs}=15^\circ$ \citep{2005AJ....130.1418J,2019ApJ...874...43L}, the corresponding Doppler factor is $\delta_{\rm quies}\approx3.5$. In external-Compton models, the observed luminosity scales strongly with Doppler boosting, so reproducing the observed Compton dominance would require a substantial increase in $\delta$ relative to this quiescent value.

However, neither varying $\Gamma$ alone nor varying $\theta$ alone provides a plausible solution. At fixed viewing angle, increasing $\Gamma$ also changes $\beta$, and hence $\delta=[\Gamma(1-\beta\cos\theta)]^{-1}$ nonlinearly; for $\theta=15^\circ$, this does not produce the required enhancement in observed luminosity. Conversely, if one keeps $\Gamma=6$ fixed and changes only $\theta$, the maximum possible Doppler factor is $\delta_{\max}\approx 2\Gamma \approx 12$, attained only for $\theta\rightarrow0^\circ$, still well below the value required to account for the observed flare purely through geometric boosting. Achieving the necessary amplification would therefore require both a substantially larger Lorentz factor and an implausibly small viewing angle within a short timescale. We therefore rule out geometric Doppler boosting as the explanation for the 2018 orphan outburst. Moreover, Doppler-factor variations are expected to produce largely achromatic variability, in contrast to the observed behavior, where a strong $\gamma$-ray flare is not accompanied by corresponding variability in the X-ray or optical bands. This further disfavors a geometric origin for the 2018 event.

\subsubsection{Magnetic reconnection in current-driven instabilities}
\label{sec:reconnection}

Magnetic reconnection driven by current-driven instabilities (CDI; \citealt{2014ApJ...783L..21S,2019ApJ...884...57B}) predicts particle acceleration in reconnection layers ($\gamma_{e}\sim10^{4}$-$10^{5}$) with strong synchrotron emission due to high particle densities and magnetic fields $B\sim$0.1-1\,G. However, this scenario fails to explain our observed Compton dominance of $L_\gamma/L_{\rm syn}\sim$160. Reproducing this ratio within the Ring of Fire scenario, i.e., for the seed photon energy density derived from the C3 region (Section~\ref{sec:ring_of_fire}), requires $B_{\rm blob}\sim$0.023\,G, which is inconsistent with the magnetization $\sigma\sim$1-10 expected in CDI models \citep{2011MNRAS.413..333N,2013MNRAS.431..355G,2014ApJ...783L..21S}.

Moreover, CDI is expected to produce more spatially extended emission distributed across the jet cross-section, whereas our VLBI observations show localized, sequential brightening as N propagates at $\mu=1.3$\,mas\,yr$^{-1}$ through features C1, C2, and C3 (Table~\ref{tab:model_fit}), indicating discrete interaction sites. Additionally, turbulent reconnection generates randomized EVPA behavior \citep{2014ApJ...780...87M}, contradicting the ordered polarization evolution seen in our data: a systematic $\Delta\chi=$24$^\circ$ rotation in C1 and a polarization peak of m=16\% in C3 (Table~\ref{tab:pol_components}). The CDI model therefore cannot account for the localized morphology, extreme Compton dominance, or ordered magnetic field structure observed during the 2018 outburst.

\subsubsection{Blob-star collision model}
\label{sec:blob_star}

\citet{2016MNRAS.463L..26B} propose that relativistic blobs can encounter luminous stars embedded within the jet, producing orphan $\gamma$-rays through Comptonization of stellar radiation. At the location of C3 in 3C~120 ($r\sim$0.38\,mas, or $\sim$0.25\,pc deprojected), this scenario would require a massive star ($\sim20$\,M$_\odot$) positioned within the jet flow.

This model predicts a brief, single-point interaction producing a symmetric $\gamma$-ray light curve. However, our VLBI observations reveal sequential brightening across three spatially distinct features, C1, C2, and C3, over $\sim$90\,days (Table~\ref{tab:model_fit}), consistent with N's continuous propagation rather than a localized stellar encounter. Furthermore, the relevant photon energy density is determined by the distance between the blob and the star, not by the distance from the central engine. For a typical luminous star with $L_\star\sim10^{37}$\,erg\,s$^{-1}$, the resulting photon energy density at plausible interaction distances is $u_\star\sim10^{-8}$\,erg\,cm$^{-3}$, significantly lower than the values required to power the observed $\gamma$-ray luminosity.

While the characteristic electron Lorentz factors are set by the energy of the target photons, the low photon energy density implies that the inverse-Compton luminosity would be severely limited. Reproducing $L_\gamma=3.7\times10^{44}$\,erg\,s$^{-1}$ would therefore require either unrealistically high particle densities or extreme bulk Lorentz factors, both inconsistent with observational constraints. Additionally, blob-star interactions predict negligible radio variability, in direct conflict with the substantial flux enhancements observed in C1-C3 during the $\gamma$-ray outburst. The combination of extended timescale ($\sim$100-150\,days), sequential spatial morphology, and tight radio/$\gamma$-ray temporal coincidence cannot be reconciled with discrete stellar collisions. We therefore rule out this mechanism for the 2018 event.

\subsection{Implications for neutrino production in a structured spine-sheath jet}

The association of the 2018 March $\gamma$-ray outburst in 3C\,120 with the IceCube neutrino event IC-180213A has recently been proposed by \citet{2025A&A...702A.129C}, who identified this outburst as the only significant $\gamma$-ray outburst temporally coincident with the neutrino arrival within a half-year window. Notably, the $\gamma$-ray outburst discussed by \citet{2025A&A...702A.129C} is the same event analysed in this work, which we establish as a clear orphan $\gamma$-ray outburst, lacking any contemporaneous variability in the optical and X-ray bands.

Our VLBI analysis provides an additional and independent constraint on the physical nature of this event. The delayed emergence of a new parsec-scale disturbance, inferred from the sequential brightening and polarization changes of the inner stationary jet components, indicates that the $\gamma$-ray outburst was associated with a dynamical restructuring of the jet occurring upstream of the 43\,GHz VLBI core. This behaviour is
naturally accommodated in models invoking a longitudinally or transversely structured jet, composed of a fast relativistic spine surrounded by a slower sheath or layer.

In such a spine-sheath configuration, the relative bulk motion between the two jet components leads to a substantial enhancement of the photon energy density in the comoving frame of the spine \citep[e.g.][]{2005A&A...432..401G,2016A&A...585A..25T}. While this effect has primarily been explored in the context of leptonic inverse-Compton emission, it also provides favourable conditions for efficient photohadronic interactions. Protons accelerated in the fast spine may interact with the amplified radiation field produced in the sheath, leading to pion production and, consequently, high-energy neutrinos, without necessarily producing a strong electromagnetic counterpart at lower energies \citep[e.g.,][]{2025ApJ...989..208P}.

Within this framework, the temporal coincidence between the $\gamma$-ray flare and the neutrino event IC-180213A may indicate a physical association \citep{2025A&A...702A.129C}, but it does not uniquely determine the radiation mechanism responsible for the observed $\gamma$-ray emission. In our interpretation, the observed $\gamma$-ray flare is most naturally explained by the leptonic Ring of Fire scenario discussed above.

Hadronic interactions may still occur in the same disturbance, or in nearby regions of the structured jet, and could contribute to neutrino production without dominating the observed electromagnetic output. We note, however, that efficient $p\gamma$ interactions are generally expected to initiate electromagnetic cascades, which would enhance the X-ray to soft $\gamma$-ray emission. The absence of such activity in our data therefore suggests that any hadronic contribution to the observed $\gamma$-ray flare must be sub-dominant or radiatively suppressed.

The structured-jet scenario therefore provides a unified explanation for
(i) the orphan character of the $\gamma$-ray outburst, (ii) the subsequent VLBI jet evolution observed in this work, and (iii) the possible production of a $\sim$ 0.1\,PeV neutrino as suggested by \citet{2025A&A...702A.129C}. While our data do not allow us to directly constrain the neutrino production efficiency, the spatial and temporal coincidence between the $\gamma$-ray outburst and the onset of a new jet disturbance supports the idea that transient dissipation events in structured jets of radio galaxies can act as multi-messenger sources.

Future coordinated campaigns combining high-cadence VLBI monitoring, $\gamma$-ray observations, and real-time neutrino alerts will be crucial to test this scenario quantitatively. In particular, detecting repeated orphan $\gamma$-ray flares accompanied by delayed VLBI signatures in nearby radio galaxies like 3C\,120 would provide strong evidence that structured jets play a key role in the production of astrophysical neutrinos.

\subsection{Comparison with previous orphan flares}
\label{sec:comparison}

The 2018 outburst differs significantly from the 2014-2015 sub-day events analyzed by \citet{2016MNRAS.458.2360J}. Those rapid $\gamma$-ray flares (doubling timescales $\sim$2-12\,h) were interpreted as fast spine reorientation ($\Gamma_\mathrm{spine}\sim$20-40, $\Delta\theta\sim1^\circ$-$3^\circ$) producing strong Doppler boosting of external Compton emission on BLR photons at $r\lesssim$0.05\,mas, deep within the BLR.

Our 2018 event spans $\sim$100-150\,days (MJD 58148-58298, epochs 2018.0-2018.4), with five-epoch VLBI coverage revealing sequential brightening of C1, C2, and C3, tracing N's propagation at $\mu=1.3$\,mas\,yr$^{-1}$ through resolved jet structure. The extreme Compton dominance ($L_\gamma/L_{\rm syn,blob}\sim$160) requires inverse-Compton scattering of C3 synchrotron photons with magnetic field $B_{\rm blob}=0.023$\,G, not geometric Doppler boosting. The emission site at $\sim$0.25\,pc is $\sim$10$\times$ beyond the BLR, ruling out BLR photons as the dominant seed field.

The 2014-2015 flares occurred close to the VLBI core where rapid geometric changes and BLR photon dominance are plausible, while the 2018 outburst arose from a well-resolved interaction at larger distances. These distinct scenarios demonstrate that 3C~120 produces orphan flares via multiple mechanisms depending on emission region location and environment.

\section{Conclusions}
\label{sec:conclusions}

We have presented a multi-wavelength study of the brightest $\gamma$-ray outburst ever observed from the radio galaxy 3C~120, combining high-cadence 43\,GHz VLBI observations with comprehensive X-ray and $\gamma$-ray monitoring. Our findings are:

\begin{enumerate}
\item The March 2018 outburst represents a clear orphan $\gamma$-ray event, with peak luminosity $L_\gamma = 3.7 \times 10^{44}$\,erg\,s$^{-1}$ but no detectable X-ray variability across the 0.3-200\,keV band, establishing extreme Compton dominance ($L_\gamma / L_{\rm syn,blob} \approx 160$).

\item VLBI imaging reveals a new disturbance (N) emerging from the core at epoch $t_{\rm ej} = (2018.01\pm0.08)$, propagating at $\mu = (1.30\pm0.57)$\,mas\,yr$^{-1}$ ($\beta_{\rm app} = (2.8 \pm 1.3)$) through the inner jet's quasi-stationary structure (C1-C3).

\item The $\gamma$-ray peak (MJD 58208, epoch 2018.2) coincides spatially and temporally with N's passage through component C3 at $r\sim$ 0.38\,mas, where we observe maximum polarization ($m = 16\%$) and significant EVPA rotation ($\Delta\chi$ $\sim$ 24$^\circ$), indicating localized magnetic field compression.

\item Analytic estimates support the Ring of Fire scenario \citep{2015ApJ...804..111M}, in which N ($\Gamma_{\rm blob} = 6$, $B_{\rm blob} = 0.023$\,G) inverse-Compton scatters synchrotron photons from C3, reproducing the observed $\gamma$-ray luminosity to within the uncertainties of the analytic estimate. Alternative models are ruled out: spine-sheath geometric boosting requires implausibly large viewing-angle swings ($\Delta\theta$ $\sim$ 14$^\circ$), magnetic reconnection predicts excessive synchrotron emission, and blob-star collisions cannot explain the extended ($\sim$100-150\,day) timescale and sequential brightening pattern.

\item The orphan outburst's association with a structured jet disturbance supports models linking VLBI-resolved features to high-energy emission \citep{2025A&A...702A.129C}, suggesting that transient interactions in mildly misaligned jets can produce enhanced photon fields for multi-messenger emission.
\end{enumerate}

The contrast with the 2014-2015 sub-day flares in the same source \citep{2016MNRAS.458.2360J} shows that structured jets can produce orphan emission through fundamentally different mechanisms depending on the interaction site, a diversity that future coordinated campaigns combining real-time neutrino alerts with high-cadence VLBI monitoring are uniquely positioned to explore.

\begin{acknowledgements}
We thank the anonymous referee for their careful reading of the manuscript and for comments that significantly improved the paper. We thank N. MacDonald and E. Goreth for the useful discussions and their insights into the Ring of Fire modeling framework.
The research at Boston University was supported in part by NASA Fermi GI grant 80NSSC20K1567.
The research leading to these results has received funding from the European Union’s Horizon 2020 Programme under the AHEAD2020 project (grant agreement n. 871158)
GB acknowledges financial support for the GRACE project, selected via the Open Space Innovation Platform (\url{https://ideas.esa.int}) as a Co-Sponsored Research Agreement and carried out under the Discovery programme of, and funded by, the European Space Agency (agreement No.
4000142106/23/NL/MGu/my).
GB and JR thank the Italian Space Agency for the financial support under the “INTEGRAL ASI-INAF” agreement n◦ 2019-35-HH.0.  The research leading to these results has received funding from the European Union’s Horizon 2020 Programme under the AHEAD2020 project (grant agreement n. 871158).
GB acknowledges financial support from the Bando Ricerca Fondamentale INAF 2023 for the project: \textit{\lq\lq The GRACE project: high-energy giant radio galaxies and their duty cycle\rq\rq}. 
The research at Boston University was supported in part by the National Science Foundation grant AST-2108622,  and several NASA Fermi Guest Investigator grants, the latest are 80NSSC23K1508 and 80NSSC23K1507. 
IA acknowledges financial support from the Spanish "Ministerio de Ciencia e Innovaci\'{o}n" (MCIN/AEI/ 10.13039/501100011033) through the Center of Excellence Severo Ochoa award for the Instituto de Astrof\'{i}isica de Andaluc\'{i}a-CSIC (CEX2021-001131-S). Acquisition and reduction of the MAPCAT data was supported by MICINN grants PID2019-107847RB-C44 and PID2022-139117NB-C44.
This study was based in part on observations conducted using the 1.8m Perkins Telescope Observatory (PTO) in Arizona, which is owned and operated by Boston University. The VLBA is an instrument of the National Radio Astronomy Observatory. The National Radio Astronomy Observatory is a facility of the National Science Foundation operated by Associated Universities, Inc.
The MAPCAT observations were carried out at the Calar Alto Observatory, which is jointly operated by Junta de Andaluc\'ia and Consejo Superior de Investigaciones Cient\'ificas.
\end{acknowledgements}

\bibliographystyle{aa} % or try abbrvnat or unsrtnat 
\bibliography{aanda}
%-------------------------------------------------------------------

\end{document}